\documentclass[aps,prl,twocolumn,showpacs,superscriptaddress,groupedaddress]{revtex4}  

\usepackage{graphicx}
\usepackage{dcolumn}
\usepackage{bm}
\usepackage{xcolor}

\usepackage{amsmath,amssymb}
\usepackage{mathtools}

\newcommand{\bea}{\begin{eqnarray}}
\newcommand{\eea}{\end{eqnarray}}
\newcommand{\simgt}{\hbox{ \raise3pt\hbox to 0pt{$>$}\raise-3pt\hbox{$\sim$} }}
\newcommand{\simlt}{\hbox{ \raise3pt\hbox to 0pt{$<$}\raise-3pt\hbox{$\sim$} }}

\newcommand{\be}{\begin{equation}}
\newcommand{\ee}{\end{equation}}

\begin{document}

\begin{flushright}
TU--1288
\end{flushright}

\title{Method to measure quarkonium wave function using $B_c$ decay}

\author{Yukinari Sumino}
 \affiliation{Department of Physics, Tohoku University,
Sendai, 980--8578 Japan
}


\date{\today}

\begin{abstract}
We propose a method that enables a direct experimental probe of the quarkonium wave function
defined in potential nonrelativistic QCD (pNRQCD) using the three-body decay of the $B_c$ meson.
We show that the momentum distribution of the spectator 
$c$-quark in the partonic decay is proportional to the absolute square of the momentum-space wave function of the $B_c$ state.
It would provide, for the first time, an experimentally accessible probe of the quarkonium wave function.
\end{abstract}

\maketitle



Heavy quarkonium states have long provided a clean testing ground for QCD dynamics \cite{QuarkoniumWorkingGroup:2004kpm,Brambilla:2010cs,Sumino:2014qpa}.
Potential-Nonrelativistic QCD (pNRQCD) \cite{Pineda:1997bj,Brambilla:2004jw}
is an effective field theory for the heavy quarkonium system, 
in which binding dynamics of quarkonium is dictated as a nonrelativistic quantum mechanical system.
In fact, the quarkonium wave function is given by a solution to the Schr\"odinger equation of quantum mechanics.

Various quarkonium observables are expressed by matrix elements defined in terms of the quarkonium wave functions, and were 
tested against the corresponding experimental data.
A direct measurement of the wave function, however, has not been achieved so far.
Such a measurement would offer a unique opportunity to directly relate experimental observables to the wave function obtained from the Schr\"odinger equation, thereby providing a clear test of our understanding of quarkonium formation dynamics.

In this paper we propose a method to measure the absolute-square of the quarkonium wave function in momentum space
using decay processes of $B_c$.
The idea is an extension of the method to measure the toponium wave function in future $e^+e^-$ collider analyses.
It was shown \cite{Sumino:1992ai,Fujii:1993mk}
that in the leading-order (LO) of the non-relativistic expansion,
the momentum distribution of the top quark in the $t\bar{t}$ threshold region is proportional to the $S$-wave Geen
function of the Schr\"odinger equation,
\bea
\left.
\frac{d\sigma}{dp_t} 
\right|_{\rm LO}
\propto
p_t^2
\left| \tilde{G}(p_t;E+i\Gamma_t) \right|^2 \,,
\\
 \tilde{G}(p_t;E+i\Gamma_t)  = \sum_n \frac{\phi_n(p_t)\psi_n(0)^*}{E-E_n+i\Gamma_t} \,,
\eea
where the top-quark momentum $\vec{p}_t $ is reconstructed from the sum of the momenta of its decay products
in the $t\bar{t}$ c.m.\ frame, and $p_t=|\vec{p}_t|$.
Experimentally $\vec{p}_t$ can be reconstructed from the momenta of the decay daughter particles in the top-quark 
hadronic decay
$t\to b\, q \, \bar{q}'$.
At the energy close to the $1S$ peak position, $E\simeq E_{1S}$, separated from other resonance states,
the Green function is dominated by the $1S$ state,
\bea
\tilde{G}(p_t,E_{1S}+i\Gamma_t) \simeq \frac{\phi_{1S}(p_t)\,\psi_{1S}(0)^*}{i\Gamma_t}
\,, 
\eea
where the wave functions for the energy eigenstates are defined by
\bea
\phi_{n}(p)= \int d^3\vec{r}\, e^{i\vec{p}\cdot\vec{r}} \psi_{n}(r)
\,, \\
\left[ -\frac{\nabla^2}{2m_r} + V(r) \right] \psi_{n}(r) = E_{n}\,\psi_{n}(r)
\label{SchrEq}
\,.
\eea
Here, $m_r=m_t/2$ is the reduced pole mass of the $t\bar{t}$ system.
Hence, $d\sigma/dp_t \simeq  {\cal N} p_t^2|\phi_{1S}(p_t)|^2$ with a $p_t$-independent constant ${\cal N}$.

Since $B_c$ decays by electroweak interaction, its decay particles are also expected to carry information of the momentum distribution
of $b$ or $c$ inside the bound state.
In particular, the LHCb experiment has recently reported observations of three-body decay modes
such as $B_c \to D\,K\,\pi$ \cite{LHCb:2025qcs}, which would enable probe of the momentum distribution,
unlike two-body decays where the magnitude of the momenta of the final-state hadrons are fixed
by kinematics.

In comparison to the top-quark decay, the picture that the final state hadrons carry the parton-level momenta
(local parton-hadron duality picture) would not be very accurate.
Nevertheless, since theoretical calculation of the three-body decays is anyway fairly difficult at present, it would be useful to compare
the parton-level theoretical predictions with the experimental data.

\begin{figure}
\begin{center}
\includegraphics[width=6cm]{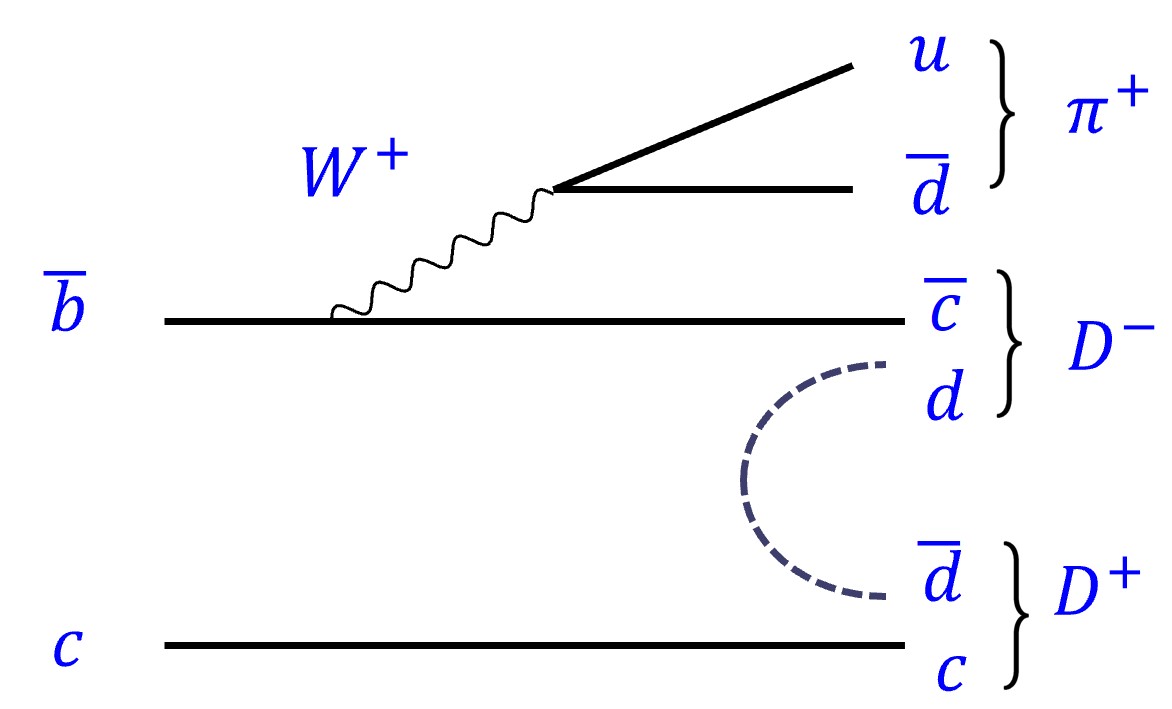}
\\
{(a)~~~~~~}\\
\includegraphics[width=6cm]{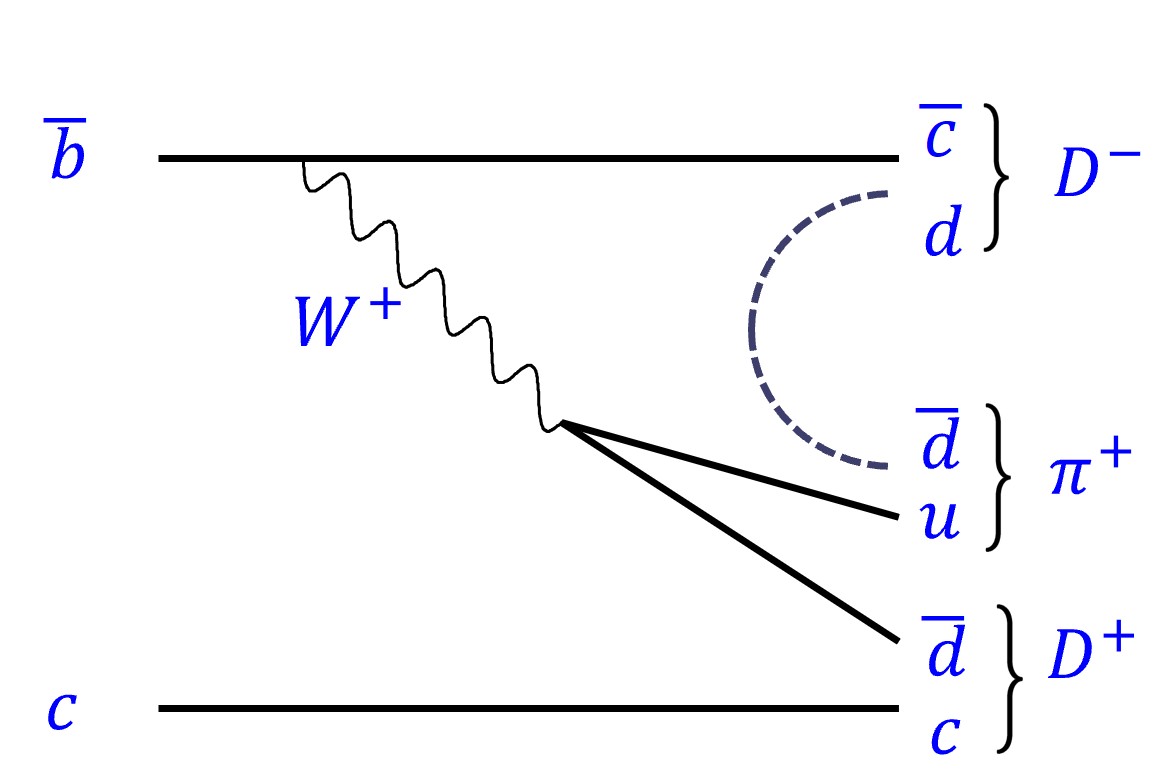}
\\
{(b)~~~~~~}\\
\includegraphics[width=6cm]{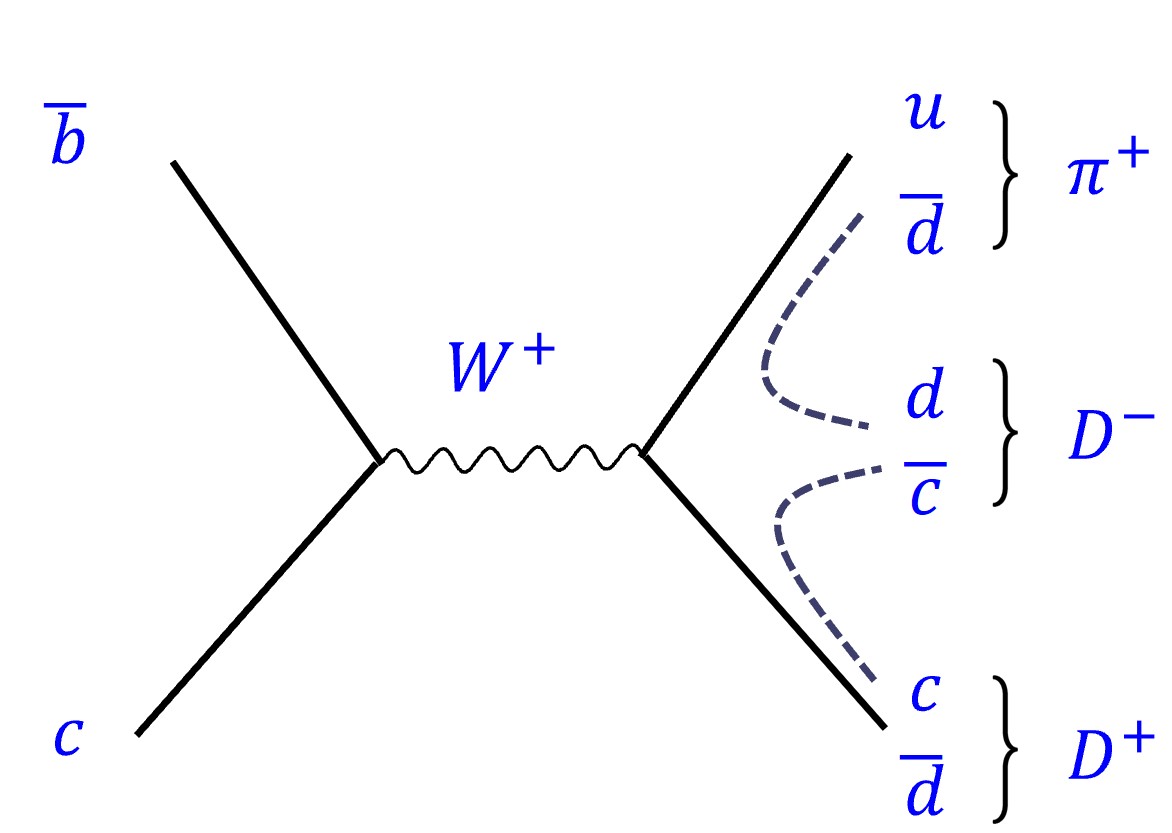}
\\
{(c)~~~~~~}
\end{center}
\caption{Feynman diagrams relevant for the process $B_c^+ \to D^+\, D^-\, \pi^+$.}
\label{fig:FeynDiag}
\end{figure}

Let us consider the decay mode $B_c^+ \to D^+\, D^-\, \pi^+$ as an example.
The dominant partonic contribution to this process is expected to be given by Fig.~\ref{fig:FeynDiag}(a),
in which the emitted color-singlet $W^+$ boson hadronizes into $\pi^+$.
In this configuration the $c$-quark acts as a spectator and carries essentially the same momentum as inside the bound state.
There are other contributions such as Fig.~\ref{fig:FeynDiag}(b) and (c).
Fig.~\ref{fig:FeynDiag}(b) would be suppressed in the kinematical region
where $c$-quark receives a large momentum transfer from $W$, 
because multi-hadron final states become increasingly favored. 
Although a fully quantitative estimate of such hadronization effects is not available, this hierarchy is consistent with the typical size of color-suppressed amplitudes.
The annihilation channel contribution
[Fig.~\ref{fig:FeynDiag}(c)] would be even more suppressed,
since the rate for the annihilation is small and production of $c\bar{c}$ in the
final state would not occur frequently.
From the total charge of the final state the parent particle can be identified as $B_c^+$,
hence, dominant part of the momentum of $\bar{b}$-quark ($c$-quark) can be reconstructed as that
of the $D^-\,\pi^+$ system ($D^+$) 
within this approximation, which would be 
sufficient for identifying the dominant contribution to the spectator momentum spectrum.

\begin{figure}
\begin{center}
\includegraphics[width=5cm]{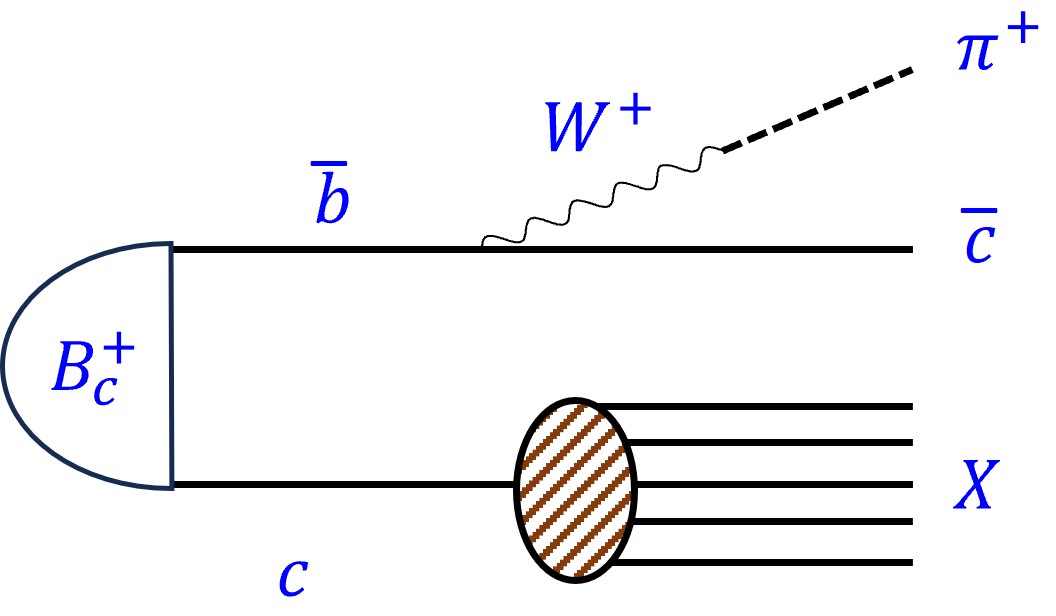}
\end{center}
\caption{
Diagram for the partonic decay amplitude eq.~\eqref{DecayAmp}.
}
\label{fig:Bc-partonic-decay}
\end{figure}

The LO partonic amplitude corresponding to Fig.~\ref{fig:Bc-partonic-decay} can be written in the $B_c^+(1S)$ rest frame as
\bea
&&
iM= \phi_{1S}(\vec{p})
\Biggl(
\frac{i}{p_c^0-m_c-\vec{p}^2/(2m_c)+i\Gamma_c/2}
\nonumber\\&&~~~
~~~~~~~~~~~~~
+
\frac{i}{p_{\bar{b}}^0-m_b-\vec{p}^2/(2m_b)+i\Gamma_b/2}
\Biggr)
\nonumber\\
&&~~~
\times  \frac{G_F}{\sqrt{2}} \times
\left(
\bar{u}_f \gamma_X \frac{1+\gamma^0}{2}\,\gamma_5 \,\frac{1-\gamma^0}{2} \gamma^\mu 
(1-\gamma_5) v_{\bar{c}}
\right)
\nonumber\\&&~~~
\times (-i) f_\pi k_\mu
\,,
\label{DecayAmp}
\eea
where
$p_{\bar{b}}^0+p_c^0=M_{B_c}$; $\vec{p}$ and $-\vec{p}$ denote the $\bar{b}$-quark and 
$c$-quark momenta, respectively;
$k$ denotes the $\pi^+$ momentum.
$\gamma_X$ and $\bar{u}_f$ depend on the decay process of $c$-quark.
For instance, in the case of the semi-leptonic decay $c \to s\,\ell^+ \,\nu$, 
\bea
\bar{u}_f \gamma_X = \frac{G_F}{\sqrt{2}} \times (\bar{\nu}\gamma^\sigma (1-\gamma_5) \ell)\times 
\bar{u}_s \gamma_\sigma (1-\gamma_5) 
\,.
\eea

The two terms in the above amplitude $iM$ 
correspond to the two possible time orderings: the decay of the $\bar{b}$
 quark or of the $c$ quark occurring first.
The first
term (with $c$ propagator) represents the former ordering,
while the second term (with $\bar{b}$ propagator) represents the latter ordering.
This form of the amplitude is not an artifact of the non-relativistic approximation but
is a general form of the partonic amplitude \cite{Lurie:1965,Sumino:2010bv}.
The $\bar{b}$-quark and $c$-quark cannot become simultaneously on-shell due to the 
binding energy.
Hence, the integral of $|M|^2$ over the whole phase space essentially decomposes into the
contributions where $c$-quark is on-shell and $\bar{b}$-quark is off-shell and vice versa.
The process $B_c^+ \to D^+\, D^-\, \pi^+$ corresponds to the former contribution.\footnote{
We assume that the pole mass $m_b$ or $m_c$ roughly approximates the constiuent quark mass
of the $b$- or $c$-quark
inside $B_c$, $B$ or $D$.
}

We take the $|1/(p_c^0-m_c-\vec{p}^2/(2m_c)+i\Gamma_c/2)|^2$ term and
integrate over the final-state phase space and over $p_b^0$, fixing $\vec{p}$.
We obtain
\bea
\left.
\frac{d\Gamma(B_c^+(1S) \to c\,\bar{c}\,\pi^+)}{d^3\vec{p}} \right|_{\rm LO}
\propto ~ \left| \phi_{1S}(\vec{p}) \right|^2
\,,
\eea
up to a $\vec{p}$-independent normalization factor.
This form is irrespective of the decay process of $c$-quark.
The same form also holds for an excited state of $B_c$.

\begin{figure}[h]
\begin{center}
\includegraphics[width=6cm]{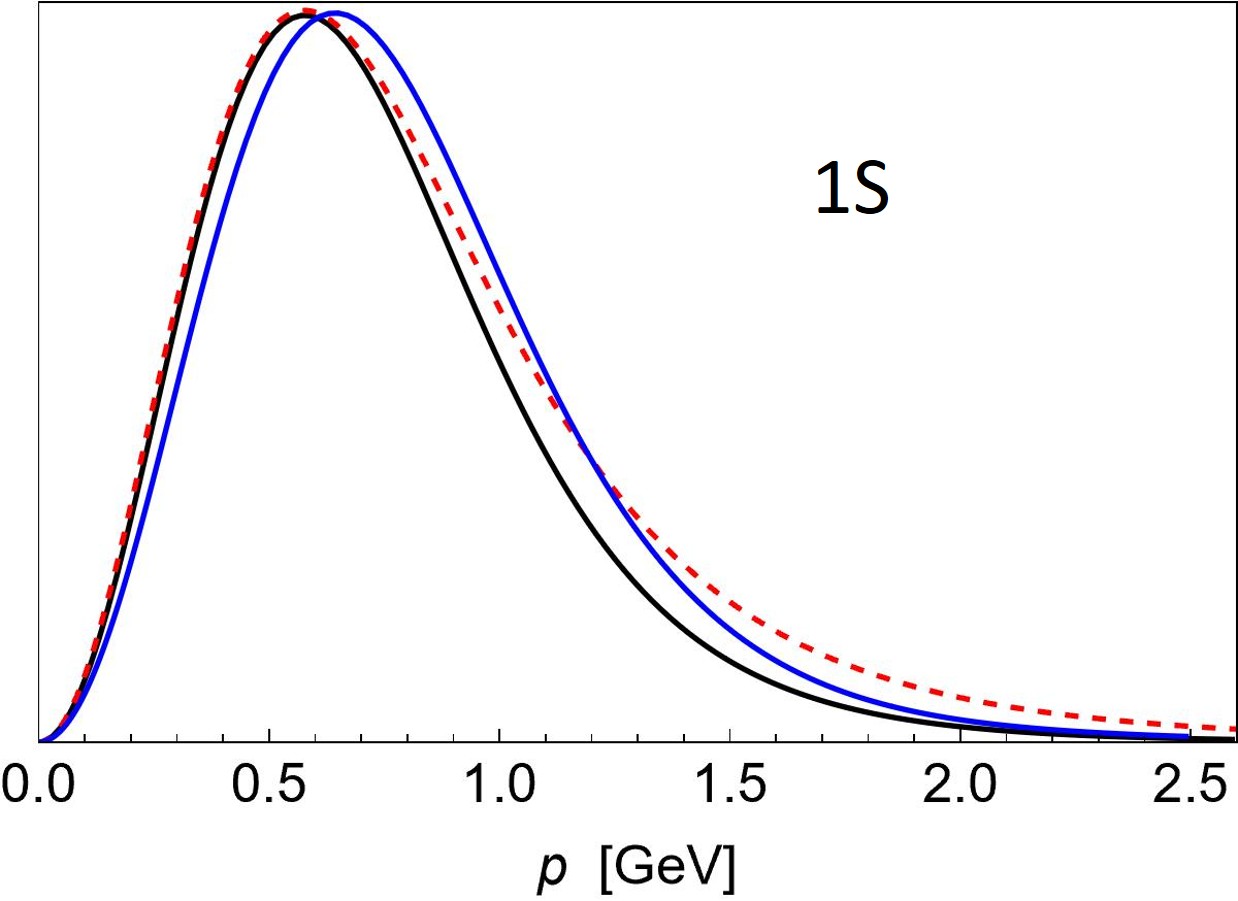}
\vspace*{5mm}\\ 
\includegraphics[width=6cm]{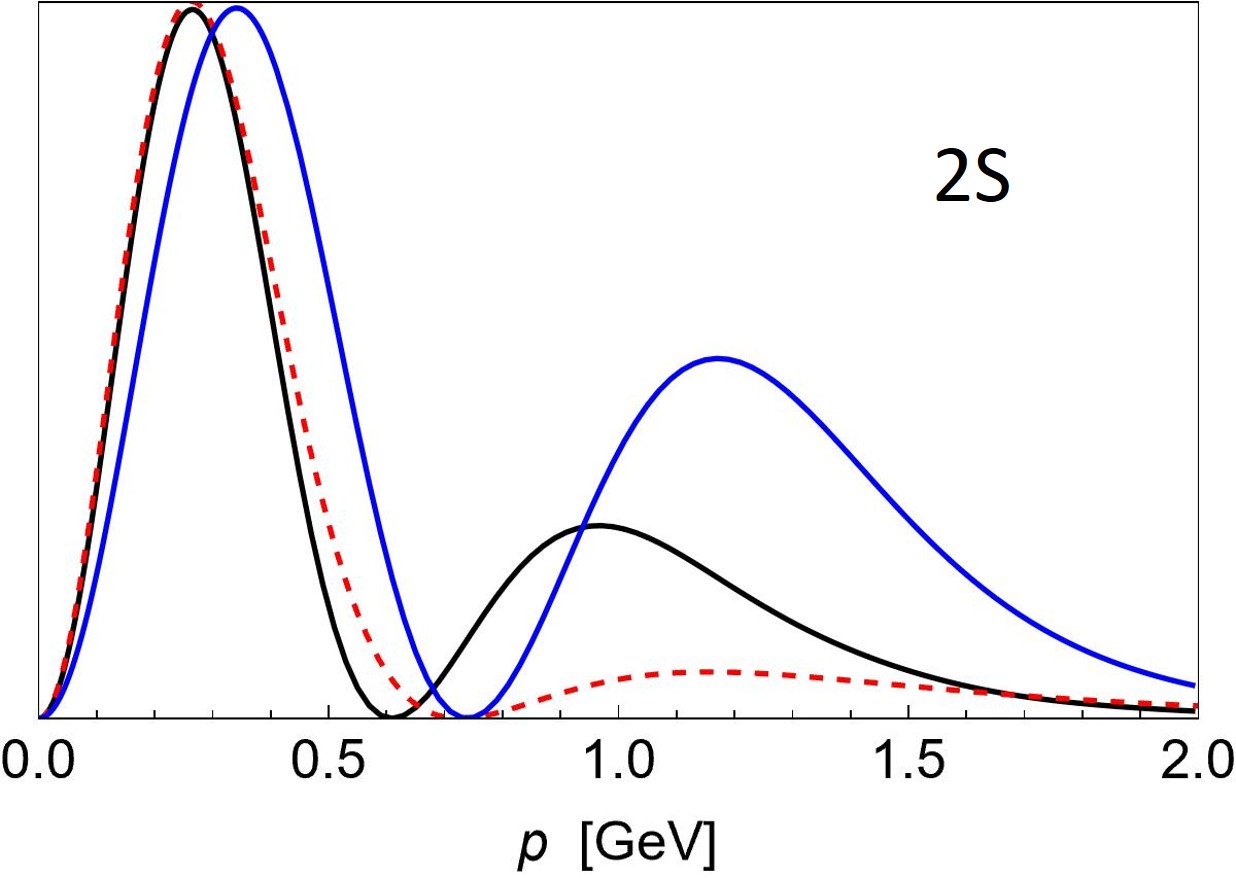}
\end{center}
\caption{
Absolute square of the momentum-space wave functions of the $1S$ and $2S$ $B_c$ states multiplied by
the phase-space factor, $p^2 |\phi(p)|^2$.
Black solid, blue solid, and red dashed lines, respectively, represent the wave functions calculated with 
the three-loop static QCD potential, Cornell potential, and Coulomb potential.
The height of $p^2|\phi(p)|^2$ is normalized at the first peak position.
}
\label{fig:Bc-WaveFn}
\end{figure}

In Fig.~\ref{fig:Bc-WaveFn} we plot $p^2\cdot|\phi(\vec{p})|^2$ for the $1S$ and $2S$
states of $B_c$
by solving the Schr\"odinger equation~\eqref{SchrEq}
with  $m_b=5.0$~GeV, $m_c=1.5$~GeV.
The solid black lines show the case where we set $V$ to the
three-loop static QCD potential \cite{Anzai:2009tm,Smirnov:2009fh}
for $\alpha_s(M_Z)=0.118$ and $n_l=3$.
For comparison, we plot $p^2\cdot|\phi(\vec{p})|^2$ in red dashed lines
in the case $V$ is chosen as the Coulomb potential, where we adjusted the
coupling strength such that the first peak positions coincide with
the QCD potential cases.
We also plot the wave functions for the Cornell potential \cite{Eichten:1978tg}
$V=-0.47/r+(0.19\,{\rm GeV}^2)r$ in blue solid lines.
All the wave functions are
normalized at the first peak.

It would not be easy to distinguish the $1S$ wave functions for the three potentials
from the experimental data.
The similarity among the $1S$ wave functions shows a major sensitivity to the Coulombic
short-distance behavior of the potentials, which can be predicted reliably by perturbative
QCD.
On the other hand, for the $2S$ wave functions, the relative magnitude of the first and second
peaks varies among different potentials.
This feature provides a potential discriminator of the underlying quarkonium dynamics.

There are a number of higher-order corrections to the above LO picture.
Concerning the dynamics of the $B_c$ bound state formation,
apart from the effects already included in the three-loop QCD potential,
all the corrections are $O(v^2)\sim O(\alpha_s^2)$ or beyond according to the
weak coupling analysis of pNRQCD, where $v$ denotes the
velocity of $b$ or $c$ quarks.
Assuming that $\alpha_s \sim 0.35$--0.5 at relevant scales, we may expect that the higher-order corrections
of order 10--25\%.
For additional non-perturbative corrections, we may regard the difference of the
wave functions between the QCD and Cornell potential cases to be an estimate of such an effect.
Approximate size of the higher-order perturbative corrections to the decay process of $B_c$ may be estimated,
for instance,
from those for the semileptonic $B$ decay
\cite{Melnikov:2008qs}.
Various moments of lepton energy or hadronic energy receive order 10\% perturbartive corrections or less.
Hence, we may expect perturbative higher-order corrections of similar size.
A systematic treatment of these effects is possible within the pNRQCD framework, but goes beyond the scope of this exploratory study.

It is also important to investigate how hadronization effects in the final states of the decay
process modify the partonic prediction.
As a crude estimate, we examine a smearing of the partonic amplitude (including the
phase) as
\bea
\tilde{\phi}(p) = \int dq\, \phi(q) f(p-q;a,\Lambda)
\,,
\\
f(\Delta p;a,\Lambda) \propto \exp \left[ -\frac{1+i a}{2}\left(\frac{\Delta p}{\Lambda}\right)^2\right]
\,,
\eea
in the range $300~{\rm MeV}\leq \Lambda \leq 500~{\rm MeV}$ and $0\leq a \leq 2$.
It shows that the second peak of the $2S$ wave function remains visible in the QCD potential case.
This may indicate that the second peak is still visible after inclusion of hadronization effects.
The smearing function $f(\Delta p;a,\Lambda)$
is introduced only as a phenomenological model to mimic typical hadronization scales, and is not intended as a substitute for a realistic nonperturbative description.

A quantitative assessment of the experimental momentum resolution is left for future work, but typical resolutions for charmed hadrons at LHCb (of order 
10-20~MeV in momentum magnitude) suggest that the peak structures should be well resolvable.

Although our estimates of the corrections to the LO predictions are very naive and crude,
once experimental data are available, comparison with the theoretical predictions would certainly
deepen our understanding and motivate us for more detailed and accurate analyses.

Finally, it is natural to ask if we can use the $B_c$ three-body decay modes including the $D$ meson 
which have already  been
observed at the LHCb experiment \cite{LHCb:2025qcs}.
These are the decays $B_c \to D\,K\,\pi$ and $B_c \to D_s\,K\,K$.
It is, however, expected that in these modes the spectator partonic contributions similar to
Fig.~\ref{fig:FeynDiag}(a) are strongly suppressed by the small Cabibbo-Kobayashi-Maskawa matrix element $|V_{ub}|$, and that the annihilation contributions similar to Fig.~\ref{fig:FeynDiag}(c) would be comparatively more important.\footnote{
For these modes, although there is another contribution from the penguin-type diagram which includes a
spectator $c$-quark, it is also
suppressed by a loop factor.
}
Since the information of the momentum distribution is lost in the annihilation channel,
these modes may not probe the wave function efficiently.

\section*{Acknowledgements}

The author is grateful to Y.~Jiang, Z.~Mu, T.~Rong, Y.~Wang, Y.~Zhang, and S.~Zheng
for fruitful discussion during the Quarkonium Workshop, Nov.~17-21, 2025.
This work was supported in part
by JSPS KAKENHI Grant Number JP23K03404.

\end{document}